   \newcommand{\be}[0]{\begin{equation}}
   \newcommand{\ee}[0]{\end{equation}}
   \newcommand{\ba}[0]{\begin{eqnarray}}
   \newcommand{\ea}[0]{\end{eqnarray}}    
\documentstyle[12pt]{article}
\begin{document}
\Large
\hfill\vbox{\hbox{DTP/98/54}
            \hbox{July 1998}}
\nopagebreak

\vspace{0.75cm}
\begin{center}
\LARGE
{\bf Complete renormalization group improvement of
QCD perturbation theory}
\vspace{0.6cm}
\Large

C.J. Maxwell

\vspace{0.4cm}
\large
\begin{em}
Centre for Particle Theory, University of Durham\\
South Road, Durham, DH1 3LE, England
\end{em}

\vspace{1.7cm}

\end{center}
\normalsize
\vspace{0.45cm}

\centerline{\bf Abstract}
\vspace{0.3cm}
%%%%%%%%%%%%ABSTRACT%%%%%%%%%%%%%%%%%%
We suggest that at any given order of Feynman diagram
calculation for a QCD observable {\it all} renormalization
group (RG)-predictable terms should be resummed to all-orders.
This ``complete'' RG-improvement (CORGI) serves to 
separate the perturbation series into infinite subsets
of terms which when summed are renormalization scheme
(RS)-invariant. Crucially all ultraviolet logarithms
involving the dimensionful parameter, $Q$, on which the
observable depends are resummed, thereby building the correct
$Q$-dependence. There are close connections with the
effective charge approach of Grunberg.
\newpage

The problem of the renormalization scale (scheme) dependence
of fixed-order perturbative QCD predictions continues to
frustrate attempts to make reliable determinations of
the underlying dimensional transmutation parameter of the
theory , ${\Lambda}_{QCD}$ (usually ${\Lambda}_{\overline{MS}}$
or ${\alpha}_{s}(M_Z)$ are the fitted quantities). Whilst
a number of proposals for controlling or avoiding this 
difficulty have been advanced \cite{r1,r2,r3,r4} no consensus has been
reached, with the result that in experimental fits attempts
are made to estimate an ad hoc ``renormalization scale''
uncertainty \cite{r5}. This ``theoretical error'' is typically
far larger than the actual experimental errors, and in our view
can completely mislead as to both the central value of
${\Lambda}_{\overline{MS}}$ and the likely importance of
uncalculated higher-order corrections \cite{r4,r6}.\\

In this letter we wish to propose that one should perform
a resummation to all-orders of {\it all} renormalization
group (RG)-predictable terms at each order of perturbation
theory. This procedure automatically organizes the series
into infinite subsets of terms which are separately
renormalization scheme (RS)-invariant, and crucially also,
by completely resumming ultraviolet logarithms, generates
the correct asymptotic dependence on the single dimensionful
parameter `$Q$' on which the observable depends. This
``complete'' RG-improvement has close connections with
other proposed solutions of the RS-dependence problem, in
particular with the effective charge approach of Grunberg
\cite{r2} which focusses on the $Q$-dependence of the observable.
The principle of resumming {\it all} known information can
also provide a guide for treating situations, such as moments
of structure functions and hard hadronic cross-sections ,
where one encounters factorization scheme (FS) dependence in
addition to RS-dependence \cite{r7}.\\

We begin by briefly reviewing the problem of parametrizing
RS-dependence, and define the concept of RG-predictable
terms. Consider the dimensionless QCD observable ${\cal{R}}(Q)$
, dependent on the single energy scale $Q$ (we assume massless
quarks). Without loss of generality, by raising   
to a power and scaling, we can arrange that ${\cal{R}}(Q)$
has a perturbation series of the form,
\be
{{\cal{R}}(Q)}=a+{r_1}{a^2}+{r_2}{a^3}+\ldots+{r_n}{a^{n+1}}
+\ldots\;,
\ee
where $a\equiv{\alpha_s}({\mu})/{\pi}$ 
is the RG-improved coupling.
The ${\mu}$-dependence of $a$ is governed by the
beta-function equation,
\be
\frac{{\partial}a}{{\partial}{\ln}{\mu}}   
=-b{a^2}(1+ca+{c_2}{a^2}+{\ldots}+{c_n}{a^n}+{\ldots})\;.
\ee
Here $b$ and $c$ are the first two universal terms of the
QCD beta-function
\ba
b&=&\frac{33-2{N_f}}{6}\;,\\
c&=&\frac{153-19{N_f}}{12b}\;,
\ea
with ${N_f}$ the number of active quark flavours. As
demonstrated by Stevenson \cite{r1} the renormalization scheme
may be completely labelled by the variables 
${\tau}{\equiv}b\ln({\mu}/{\tilde{\Lambda}})$ and the
non-universal beta-function coefficients ${c_2},{c_3},{\ldots}$.
$a({\tau},{c_2},{c_3},{\ldots})$ is obtained as the solution
of the transcendental equation
\be
\frac{1}{a}+c\ln\left(\frac{ca}{1+ca}\right)=
{\tau}-\int^{a}_{0}\;{dx}\;\left(-\frac{1}{B(x)}+
\frac{1}{{x^2}({1+cx})}\right)\;,
\ee
where $B(x)\equiv{x^2}(1+cx+{c_2}{x^2}+{c_3}{x^3}+{\ldots})$.
This is obtained by integrating Eq.(2) with a suitable choice
of boundary condition related to the definition of
${\tilde{\Lambda}}$.\\

For our purposes it will be more useful to label the RS
using $r_1$, the next-to-leading order (NLO) perturbative
coefficient, rather than $\tau$. This is possible because
\cite{r1}
\be
{\tau}-{r_1}={{\rho}_{0}}(Q){\equiv}b\ln(Q/{{\Lambda}_{\cal{R}}})
\;,
\ee
where ${\rho}_{0}$ is an RS-invariant, hence ${\tau}$ can be
traded for ${r_1}$. ${\Lambda}_{\cal{R}}$ is a dimensionful
scale dependent on the particular observable. It is related
to the {\it universal} dimensional transmutation
parameter ${\tilde{\Lambda}}_{\overline{MS}}$ by 
\be
{\Lambda}_{\cal{R}}{\equiv}{e}^{r/b}{\tilde{\Lambda}}_
{\overline{MS}}\;,
\ee
where $r{\equiv}{r}_{1}^{\overline{MS}}({\mu}=Q)$. The
righthand side of Eq.(7) is independent of the subtraction
scheme employed, and as we shall see ${\Lambda}_{\cal{R}}$
has a physical significance, being directly related to
the asymptotic $Q$-dependence of ${\cal{R}}(Q)$ \cite{r2,r4,r6}.
Equations (5),(6) can then be used to define 
$a({r_1},{c_2},{c_3},{\ldots})$.\\

We next turn to the RS-dependence of the perturbative
coefficients $r_i$. This must be such as to cancel the
RS-dependence of `$a$' when the series is summed to 
all-orders. The self-consistency of perturbation theory
\cite{r1} demands that the result of a ${\rm{N}}^n$LO 
calculation (terms up to and including ${r_n}{a}^{n+1}$) in
two {\it different} schemes should differ by O(${a}^{n+2}$).
This implies the following dependences of the $r_i$ on the
scheme parameters-
 ${r_2}({r_1},{c_2})$, ${r_3}({r_1},{c_2},
{c_3})$,${\ldots},{r_n}({r_1},{c_2},{c_3},{\ldots},{c_n})$.
The explicit dependences imposed by self-consistency are
found to be
\ba
{r_2}({r_1},{c_2})&=&{r_1}^{2}+c{r_1}+{X_2}-{c_2}
\nonumber\\
{r_3}({r_1},{c_2},{c_3})&=&{r_1}^{3}+{5\over2}c{r_1}^{2}
+(3{X_2}-2{c_2}){r_1}+{X_3}-{1\over2}{c_3}
\nonumber\\
\vdots & &\vdots\;.
\ea
In general the structure is
\be
{r_n}({r_1},{c_2},{\ldots},{c_n})={{\hat{r}}_{n}}({r_1},{c_2},
{\ldots},{c_{n-1}})+{X_n}-{c_n}/(n-1)\;.
\ee
Here ${\hat{r}}_{n}$ is an $n^{\rm{th}}$ order polynomial in
${r_1}$
which is determined given a complete
${\rm{N}}^{n-1}$LO calculation. $X_n$ is $Q$-independent
and RS-invariant and can only be determined given a complete
${\rm{N}}^{n}$LO calculation. ${\hat{r}}_{n}$ is the 
``RG-predictable'' part of $r_n$ , and $X_n$ is 
``RG-unpredictable''.
Thus, given a NNLO calculation in the ${\overline{MS}}$ scheme
with ${\mu}=Q$ one can determine the RS-invariant
\be
{X_2}={{r}_{2}^{\overline{MS}}}({\mu}=Q)
-{({r}_{1}^{\overline{MS}}({\mu}=Q))}^{2}
-c\;{{r}_{1}^{\overline{MS}}({\mu}=Q)}+{c}_{2}^{\overline{MS}}\;,
\ee
where
\be
{c}_{2}^{\overline{MS}}=\frac{77139-15099{N_f}+325{N}_{f}^{2}}
{1728b}\;.
\ee
By a complete ${\rm{N}}^{n}$LO calculation we mean that
${c_2},{c_3},{\ldots},{c_n}$ have been computed as well as
${r_1},{r_2},{\ldots},{r_n}$. We note in passing that
${c}_{3}^{\overline{MS}}$ has now been calculated \cite{r8}.\\

Using Eqs.(8) we can now exhibit the explicit RS-dependence of
the terms of Eq.(1),
\be
{\cal{R}}(Q)=a+{r_1}{a^2}+({r}_{1}^{2}+c{r_1}+{X_2}-{c_2}){a^3}
+({r}_{1}^{3}+{5\over2}c{r}_{1}^{2}+(3{X_2}-2{c_2}){r_1}
+{X_3}-{1\over2}{c_3}){a^4}+{\ldots}\;,
\ee
where $a{\equiv}a({r_1},{c_2},{c_3},{\ldots})$.
We now implement the idea of ``complete'' RG-improvement (CORGI).
At any given order of Feynman diagram calculation {\it all}
known (RG-predictable) terms should be resummed to all-orders.
Given a NLO calculation ${r_1}$ is known but
${X_2},{X_3},{\ldots}$ are unknown. Thus the complete subset
of known terms in Eq.(12) at NLO is
\be
{a_0}{\equiv}a+{r_1}{a^2}+({r}_{1}^{2}+c{r_1}-{c_2}){a^3}
+({r}_{1}^{3}+{5\over2}c{r}_{1}^{2}-2{c_2}{r_1}-{1\over2}{c_3})
{a^4}+{\ldots}\;.
\ee
The sum of these terms, $a_0$ , can be simply determined using the
following two-step argument. The infinite subset of terms in
Eq.(13) has an RS-independent sum, since the ${X_2},{X_3},{\ldots}$,
-dependent terms cannot cancel their RS-dependence, and we know
that the full sum of Eq.(12) is RS-invariant. Each term is
a multinomial in ${r_1},{c_2},{c_3},{\ldots}$. Using the
RS-independence we can set ${r_1}=0,{c_2}=0,{c_3}=0,{\ldots}$, in
which case all terms but the first in Eq.(13) vanish and we
obtain ${a_0}=a({r_1}=0,{c_2}=0,{c_3}=0,{\ldots},{c_n}=0
,{\ldots})$. So at NLO CORGI corresponds to working in an
```t Hooft scheme'' with ${c_2}={c_3}={\ldots}=0$ \cite{r9},
and with ${r_1}=0$. From Eq.(6) ${r_1}=0$ corresponds to
${\tau}=b\ln(Q/{{\Lambda}_{\cal{R}}})$ or to an
${\overline{MS}}$ scale ${\mu}={e}^{-r/b}Q$. This is the
so-called ``fastest apparent convergence'' (FAC) or
effective charge (EC) scale \cite{r1,r2}. 
From Eq.(5) we find that ${a_0}$ satisfies
\be
\frac{1}{a_0}+c\ln\left(\frac{c{a_0}}{1+c{a_0}}\right)
=b\ln\left(\frac{Q}{{\Lambda}_{\cal{R}}}\right)\;.
\ee
We note to avoid confusion that the definition of 
$\tilde{\Lambda}$ on which Eq.(5) is based \cite{r1} differs
from that usually used for ${\Lambda}_{\overline{MS}}$. In
terms of the standard definition we have,
\be
{\Lambda}_{\cal{R}}={e}^{r/b}{\left(\frac{2c}{b}\right)}^{-c/b}
{\Lambda}_{\overline{MS}}\;.
\ee

If a NNLO calculation has been completed , then ${X_2}$ can be
determined (as in Eq.(10)), and a further infinite subset of
terms are known and can be resummed to all-orders,
\be
{X_2}{a_0}^{3}={X_2}{a^3}+3{X_2}{r_1}{a^4}+\ldots\;.
\ee
The RS-independence of the sum and the multinomial structure
of the coefficients again leads to a resummed result involving
${a_0}$.
CORGI at each order finally leads to
\be
{\cal{R}}(Q)={a_0}+{X_2}{a_0}^{3}+{X_3}{a_0}^{4}+{\ldots}
+{X_n}{a_0}^{n+1}+{\ldots}\;,
\ee
which is simply the perturbation series in the RS with
${r_1}={c_2}={c_3}={\ldots}={c_n}={\ldots}=0$.\\

One can make a very strong case for the principle of
complete RG-improvement. Clearly we cannot resum to
all-orders {\it more} terms than are exactly known, and
the only reason to leave out some known terms would be
if their omission
compensated the effect of  higher-order
corrections $X_n$, but these contributions are simply
unknown. The ${X_n}$ can be used to parametrize uncalculated
higher-order corrections. A further crucial feature of CORGI
is that it serves to generate the correct $Q$-dependence of
the observable via the resummation of all ultraviolet
logarithms. To see this it will be useful to first briefly
review the effective charge aproach of Grunberg \cite{r2} which
is based on specifying the $Q$-dependence of ${\cal{R}}(Q)$.\\

In the effective charge (EC) approach one considers the
$Q$-evolution of ${\cal{R}}(Q)$ in terms of ${\cal{R}}(Q)$
itself,
\be
\frac{{\partial}{\cal{R}}(Q)}{{\partial}\ln{Q}}=
-b{\cal{R}}^{2}(1+c{\cal{R}}+{\rho_2}{\cal{R}}^{2}+
{\rho_3}{\cal{R}}^{3}+\ldots)
{\equiv}-b{\rho}({\cal{R}})\;.
\ee
This equation may be regarded as the beta-function equation
(Eq.(2)) in the RS where ${r_1}={r_2}=\ldots={r_n}=\ldots=0$,
so that ${\cal{R}}=a$. This effective charge scheme corresponds
to ${r_1}=0,{c_2}={\rho_2},\ldots,{c_n}={\rho_n},\ldots$
. The ${\rho_n}$ are RS-invariant and $Q$-independent
combinations of the perturbation series and beta-function
coefficients with, for instance,
\be
{\rho_2}={r_2}+{c_2}-{r_1}c-{r}_{1}^{2}\;.
\ee
These invariants are closely connected with the $X_n$ invariants 
which arise in the CORGI approach. One finds that
\be
{X_2}={\rho_2},\;{X_3}={\rho_3}/2,\;{X_4}={\rho_4}/3+2{\rho}_{2}^{2}
+c{\rho_3}/6,\;{\ldots}\;.
\ee
Integrating up Eq.(18) and applying asymptotic freedom
(${\cal{R}}(Q){\rightarrow}0$ as $Q{\rightarrow}{\infty}$)
as a boundary condition one obtains \cite{r2,r4}
\be
\frac{1}{\cal{R}}+c\ln\left(\frac{c{\cal{R}}}{1+c{\cal{R}}}
\right)=b\ln\left(\frac{Q}{{\Lambda}_{\cal{R}}}\right)-
\int_{0}^{\cal{R}}\;{dx}\;\left(-\frac{1}{{\rho}(x)}+
\frac{1}{{x^2}({1+cx})}\right)\;.
\ee
Alternatively Eq.(21) can be regarded as Eq.(5) in the EC
scheme where ${\tau}={\rho_0}=b\ln(Q/{{\Lambda}_{\cal{R}}})$,
and $a={\cal{R}}$.
From Eq.(21) we see that the leading large-$Q$ behaviour
of ${\cal{R}}(Q)$ is
\be
{\cal{R}}(Q){\approx}G\left(b\ln\frac{Q}{{\Lambda}_{\cal{R}}}
\right)={a_0}\;,
\ee
where $G(x)$ is the inverse function to $1/x+c\ln(cx/(1+cx))$.
For $c=0$ this reduces to ${\cal{R}}(Q){\approx}1/b\ln(Q/
{{\Lambda}_{\cal{R}}})$. So, as advertised, ${\Lambda}_{\cal{R}}$
has a physical significance in setting the large-$Q$ evolution
of the observable ${\cal{R}}(Q)$.\\

The crucial point which we now wish to demonstrate is that 
in the CORGI approach the above $Q$-dependence emerges only
if the $r_1$ terms are resummed to all-orders.
We first note that from Eq.(6) we may write
\be
{r_1}({\mu})=\left(b\ln\frac{{\mu}}{{\tilde{\Lambda}}_
{\overline{MS}}}-b\ln\frac{Q}{{\Lambda}_{\cal{R}}}\right)\;,
\ee
with ${\mu}$ taken to be the $\overline{MS}$ scale as is
customary. The ${\mu}$-dependent logarithm is completely
irrelevant since, as we shall demonstrate, all dependence
on ${\tau}{\equiv}b{\ln}({\mu}/{{\tilde{\Lambda}}_{\overline
{MS}}})$ disappears on resumming the $r_1$ terms to
all-orders. The correct $Q$-dependence of ${\cal{R}}(Q)$ is
generated by all-orders resummation of the ultraviolet
(UV) logarithms ${\ln}(Q/{{\Lambda}_{\cal{R}}})$. Normally
one would choose ${\mu}=xQ$ with $x=1$ the allegedly
``physical'' scale favoured in analyses, we stress however
that the correct $Q$-dependence still results even if
$\mu$ is taken to be a dimensionful constant independent
of $Q$. $\mu$ is truly an irrelevant parameter, it is the
resummation of all UV logarithms that is crucial.\\

To see the $Q$-dependence emerge it is convenient to
simplify the algebra by setting $c=0$ and
${c_2}={c_3}=\ldots=0$. Then 
\be
a({\mu})=1/b\ln({\mu}/{{\tilde{\Lambda}}_{\overline{MS}}})\;,
\ee
and the NLO RG-improvement in Eq.(13) becomes a geometrical
progression in $r_1$,
\be
{\cal{R}}(Q){\approx}a({\mu})+{r_1}({\mu}){a^2}({\mu})
+{r}_{1}^{2}({\mu})
{a}^{3}({\mu})+\ldots\:.
\ee
Substituting Eq.(23) for ${r_1}({\mu})$, summing the geometrical
progression and using Eq.(24) for $a({\mu})$ yields
\be
{\cal{R}}(Q){\approx}a({\mu})/\left[1-\left(b\ln\frac{\mu}
{{\tilde{\Lambda}}_{\overline{MS}}}-b\ln\frac{Q}{{\Lambda}_
{\cal{R}}}\right)a({\mu})\right]=1/b\ln(Q/{\Lambda}_
{\cal{R}})\;.
\ee
Since the $X_n$ are $Q$-independent these UV logarithms are
the only terms involved in generating the $Q$-dependence
of the observable, and they must be completely resummed
to all-orders. This resummation may be accomplished by using
the FAC (EC) scale ${\mu}={e}^{-r/b}Q$. The standard
practice of using the ``physical'' scale ${\mu}=Q$
effectively truncates the summation of the geometrical
progression and if the NLO coefficient $r$ is large can
introduce considerable inaccuracy into the extraction
of ${\tilde{\Lambda}}_{\overline{MS}}$.\\

Eqs.(20) for the $X_n$ and ${\rho}_n$ invariants make it
clear that the EC and CORGI approaches are closely related.
In NLO they are identical and in higher-orders correspond
to different RS-invariant ways of parametrizing the
uncalculated higher-order corrections. Crucially in both
approaches all $Q$-dependent UV logarithms are resummed
to all-orders, thus reproducing correctly  the Q-dependence
of the observable. These two approaches therefore represent
equally reasonable RG-improvements of QCD perturbation theory,
and the results for ${\tilde{\Lambda}}_{\overline{MS}}$ or
${\alpha}_{s}({M_Z})$ obtained using them should be
comparable unless the first uncalculated $X_n$ or ${\rho}_n$
invariant is large, in which case fixed-order perturbation
theory will be unreliable.
We note that the ``principle of minimal sensitivity'' (PMS)
approach \cite{r1} also results in an RG-improvement
sharing these desirable features.\\                   

We should finally comment on the operational significance
of these ideas for the extraction of ${\Lambda}_{\overline{MS}}$
(or ${\alpha}_{s}({M_Z})$) from experimental data .
In the standard approach (as reviewed in Ref.\cite{r5})
one determines ${\Lambda}_{\overline{MS}}$ by fitting
the NLO $\overline{\rm{MS}}$ scheme prediction with
scale ${\mu}=xQ$ to the data. Typically $x=1$ is
taken to provide a central value and variation between (say)
$x=0.5$ and $x=2$ provides an upper and lower theoretical
error estimate. This theoretical error is supposed to
be an estimate of the effect of ``uncalculated higher
order corrections''. Our point is that an infinite subset
of these uncalculated corrections are RG-predictable,
and can and should be resummed to all-orders, this
resummation being required to build the correct
Q-dependence of the observable. The truly RG-unpredictable
and unknown higher-order corrections are then
parametrized by the first uncalculated $X_n$ or
${\rho}_n$ RS-invariant. Using the effective charge
scale ${\mu}={e}^{-r/b}Q$ the scatter in the
${\Lambda}_{\overline{MS}}$ values obtained for
different observables at NLO then reflects the relative
differences in the size of $X_2$ or ${\rho}_2$  \cite{r4}. In
particular the hypothesis that these RG-unpredictable
NNLO effects are small is tested directly if the
values of ${\Lambda}_{\overline{MS}}$ obtained for
a range of observables are similar. This exercise
has been performed (for EC and PMS) in Ref.\cite{r10},
although we disagree with the final conclusions reached
in that work.\\  

The experimental isolation of power correction effects
for observables is also complicated by scale dependence
and should also be carried out in this framework, as has
recently been discussed in Ref.\cite{r6}. The
approach can also be straightforwardly modified
to eliminate the matching ambiguity in the resummation of
large kinematical logarithms for jet observables
\cite{r4}.\\

Whilst the EC, CORGI and PMS approaches all represent
equally well-motivated RG-improvements, CORGI should
enable somewhat more straightforward fitting to data, and
should also considerably simplify the renormalon-inspired  
RS-invariant ``leading-$b$'' all-orders resummations
of perturbation theory suggested in Ref.\cite{r11}. We
hope to report on these and other applications in future
work.

\section*{Acknowledgements}

We would like to thank John Campbell and Nigel Glover
for a productive collaboration and numerous discussions
which were instrumental in developing the ideas reported
here.

\end{document}